\documentclass[11pt,a4paper]{article}
\usepackage{a4wide}
\usepackage{amssymb}
\usepackage{amsmath}
\usepackage{graphicx}
\usepackage{wrapfig}
\usepackage{caption}
\usepackage{subcaption}
\usepackage{bbm}
\usepackage{cite}
\usepackage{hyperref}
\usepackage{soul}

\newcommand{\be}{\begin{equation}}
\newcommand{\ee}{\end{equation}}
\newcommand{\bea}{\begin{eqnarray}}
\newcommand{\eea}{\end{eqnarray}}

\newcommand{\bdm}{\begin{displaymath}}
\newcommand{\edm}{\end{displaymath}}

\begin{document}
{\renewcommand{\thefootnote}{\fnsymbol{footnote}}
\begin{titlepage}

\noindent
\begin{center}
%\vspace*{1cm}
{\huge Gravity-induced entanglement as a probe of spacetime curvature}

\vskip 3cm

{\large Suddhasattwa Brahma$^1$\footnote{Corresponding author: {\tt suddhasattwa.brahma@gmail.com}} and Abhinove Nagarajan Seenivasan$^2$\footnote{ \tt abhinove523@gmail.com}}  \\
\vspace{1cm}

\normalsize
$^1$Higgs Centre for Theoretical Physics, School of Physics \& Astronomy,
\\ University of Edinburgh, Edinburgh EH9 3FD, UK. \\ \vspace{0.2cm}
$^2$Department of Physics, Indian Institute of Technology Guwahati,\\ Guwahati 781039, Assam, India.\vspace{1cm}

\begin{abstract}
It is now widely believed that if the gravitational field is (perturbatively) quantum, it would entangle two massive objects (in spatial superpositions) which were otherwise unentangled to begin with. Recently, actual table-top experiments have been proposed to test this idea in what would be the first detection of perturbative quantum gravity. In this essay, we devise a thought experiment to prove that such gravity-induced entanglement depends on the spacetime curvature and can, in principle, act as an alternate signature of the expanding background. This will open up new and complementary directions to search for such entanglement in curved spacetime and reveal fresh perspectives on it.
\end{abstract}

\vskip 3cm

{\it Honourable Mention,} Gravity Research Foundation 2023 Awards for Essays on Gravitation\\[2mm] %Submitted on \today
\end{center}

\end{titlepage}
\setcounter{footnote}{0}
{\renewcommand{\thefootnote}{\arabic{footnote}}

Gravity-induced entanglement of massive objects is one of the oldest \textit{Gedankenexperimente} to test the quantum nature of the gravitational field \cite{DeWitt:1957obj}. Thanks to the advent of modern technology, variations of it are now being proposed as actual experiments to test perturbative quantum gravity in the laboratory \cite{Mari:2015qva,Bose:2017nin,Marletto:2017kzi}. The fundamental idea behind this is to put two massive objects in spatial superpostions for some time $\Delta T$, and then the gravitational phase difference between the different trajectories of these objects would put them in an entangled state, provided the gravitational field in fundamentally quantum \cite{Christodoulou:2022mkf}. In this essay, we put forward a simple generalization of this thought experiment -- What would be the effect of such gravity-mediated entanglement if the two bodies are placed in an expanding spacetime?

Although such entanglement of massive bodies have been considered extensively in recent times, let us recap a version of this as follows \cite{Bose:2022uxe}. Say, we put two free particles (A \& B) of masses $m_{A}$ and $m_{B}$ in two harmonic traps, separated by a distance $r$. These two quantum harmonic oscillators are stationary and, to begin with, non-interacting. The main conclusion of such an experiment is that even when the two oscillators are in an unentangled state to begin with, they will end up in an entangled state after some time due to the gravitational interaction between them, provided the gravitational field is quantum in nature. The main underlying assumptions for the above conclusion are the field theoretic nature of gravitational interaction (in particular, of relativistic locality) and the quantum information principle of ``\textit{LOCC}'' which states that classical communication between two objects cannot entangle them \cite{Marletto:2017kzi, Christodoulou:2022mkf}. 

Let us go down to a 20,000 ft view of the problem from the 50,000 ft one in order to understand the mathematics a little better. The Hamiltonian for the free oscillator system (in the absence of any gravitational perturbation) is given by: 
\begin{equation}
	{H}_0 \equiv {H}_A + {H}_B = \frac{\hat{p}_A^2}{2m_A} + \frac{1}{2}m_A\omega^2 \delta \hat{r}_A^2 + \frac{\hat{p}_B^2}{2m_B} +  \frac{1}{2}m_B \omega^2 \delta \hat{r}_B^2
\end{equation}
where $\delta r$ denotes the displacement from the trap location $r_A = -\frac{r}{2} + \delta r_A $ and $r_B = \frac{r}{2} + \delta r_B$. The idea is to turn on the gravitation field for such a system and consider the weak field regime in a Minkowski background, \textit{i.e.} the gravitational field is given by $g_{\mu\nu} = \eta_{\mu\nu} + \hat{h}_{\mu\nu}$, the hat over the gravitational field is to emphasize its quantum nature (we will drop the hats from now on). The interaction of the gravitational field with the stress-energy tensor $T^{\mu\nu}$ of the harmonic oscillators is given by (the particles are assumed to be at rest and we set $c=1$ throughout):
\begin{eqnarray}\label{HI&Tmunu}
 H_{I}^{\rm flat} = \frac{1}{2}\int d^3x \text{ } h_{\mu \nu} T^{\mu \nu}\,, \hspace{4mm} \text{with} \hspace{4mm}  T^{00}({r}, t) = \sum_{i = A,B} m_i\, \delta ({r} - {r_i(t)})\,.
\end{eqnarray}

Expanding the Einstein-Hilbert action to second order in $h_{\mu\nu}$, it is easy to solve for the gravitational field by expanding it in the usual positive and negative frequency modes, where the mode functions correspond to the usual Minkowski vacuum. One can plug the solution of the graviton field equations into the on-shell gravitational action to derive the interaction potential between the two oscillators \cite{Christodoulou:2022mkf}. The approximation of point particles, and only keeping the Gaussian part of the gravity action, is what allows one to exactly solve the field equations to derive the gravitational potential, \textit{i.e.} the non-relativistic limit of the scattering amplitude between the two particles corresponding to the exchange of off-shell gravitons, as 
\begin{eqnarray}\label{NewPot}
	V_{I}^{\rm flat} = - G \, \frac{m_A \,m_B}{|{r}_A - {r}_B|}\,.
\end{eqnarray}
This is the most crucial point of the gravity-induced entanglement. Even if the reader is unfamiliar with the how the derivation of this non-relativistic gravitational potential comes about from Einstein gravity, the main takeaway message is that the free oscillators would become entangled due to an effective potential acting on them coming from the interaction between the point masses and the graviton field \cite{Krisnanda:2019glc}. In particular, recently it has been shown that the Newtonian potential in \eqref{NewPot} is capable of acting as a quantum communication channel in order to entangle the two masses \cite{Belenchia:2019gcc}, and the spurious distinction between longitudinal and transverse modes of the graviton is irrelevant for this experiment \cite{Christodoulou:2022knr}. In practice, one expands the above interaction potential in small $\delta{r}_{A,B}/r \ll 1$ and work with a quadratic effective potential $\propto \delta{r}_A \delta{r}_B$ which is responsible for the leading-order entanglement \cite{Bose:2022uxe}. 

Our motivation to generalize this to cosmology is that the `vacuum' of QFT in curved space is typically more complicated than its Minkowski counterpart and it often leads to surprises. A recent example would be that the `noise' of gravitons, a complementary smoking gun signal of perturbative quantum gravity, is unobservably tiny for flat space vacuum and can yet become quite significant for quantum states with exponentially large `squeezing', as can only be obtained from de Sitter like expansion (or black hole geometries) \cite{Parikh:2020kfh}. This matches our physical intuition that inflation is capable of sourcing galaxy distributions in the universe starting from tiny zero-point fluctuations, thanks to radical effects of particle production in curved spacetimes. We ask if there is anything analogous for gravity-meditated entanglement when considering such background expansion even when taking the non-relativistic approximation for the two masses in question.

To put it more succinctly, we ask the following question: What happens to this experiment if we linearize gravity around some cosmological metric, \textit{i.e.} $g_{\mu\nu} = \eta_{\mu\nu}^{\rm FLRW} + h_{\mu\nu}$, where $\eta_{\mu\nu}^{\rm FLRW}$ stand for the flat FLRW metric? To understand the subtleties of this a bit better, recall that we mentioned the need to expand the graviton field in terms of the Minkowski (vacuum) mode functions, as per the textbook treatment of QFT in flat space, in order to derive \eqref{NewPot}. In general FLRW spacetimes, the definition of a vacuum is much more ambiguous. One might refer to an instantaneous vacuum, an adiabatic vacuum or something else. These are well-known difficulties of QFT in curved space, and even after one is able to pick among these choices using some physical criterion, the key point is that the mode functions, corresponding to such a curved space vacuum, will be different from the unambiguous Minkowski vacuum in flat space. This, in turn, implies that the effective gravitational potential (even in the leading-order non-relativistic limit) will be different from the Newtonian potential in \eqref{NewPot}, leading to a new term which will be responsible for modifying the entanglement profile. Another way to say the same thing would be to consider the graviton propagator, after expanding the Einstein-Hilbert action to quadratic power around FLRW spacetime, and derive the effective potential from the scattering amplitude between two  (non-relativistic) massive objects. The graviton propagator would pick up corrections due to background expansion, compared to flat space, and therefore the entangling potential would also be different from the standard Newtonian one in \eqref{NewPot}. 

Even without doing any computations, we are sure that for small separation $|r_A - r_B|$ between the two masses, the effective potential must coincide with \eqref{NewPot}. The effect of spacetime curvature would be felt only when  the distance between the two masses are large. However, as mentioned above, there are various ambiguities to overcome in order to derive the effective potential for FLRW geometries\footnote{For instance, even if we are to start with identifying some positive energy mode expansion at some initial time, cosmological phase transition would lead to that state becoming an excited one \cite{Vilenkin:1982wt}. This can give rise to interesting new effects but, at the same time, unnecessarily complicate our main argument here.}. Therefore, we will leave computing the interaction potential for general FLRW expansion for later and emphasize that the entanglement profile, due to the graviton-meditated interaction between two massive objects, depends on the vacuum state and thus on the metric of the background spacetime. We reiterate that there will be other differences when considering gravity-induced entanglement in FLRW \textit{vis-\`a-vis} flat space, such as in factors of the scale factor appearing in the analogous equations for \eqref{HI&Tmunu}; however, the main physical effect comes from the structure of the vacuum state (and its associated mode functions) for gravitons in cosmology as opposed to the standard Minkowski one.

Let us gain some intuition regarding how gravity-mediated entanglement can play a novel role in probing the curvature of spacetime from looking closely at the case of de Sitter space. Its metric can be written as (the Hubble parameter, $H$, being constant and $\tau$ denoting conformal time):
\begin{eqnarray}\label{dS}
	{\rm d}s^2 = \frac{1}{H^2\tau^2} \left(-{\rm d}\tau^2 + {\rm d}{\bf x}^2\right)\,.
\end{eqnarray}
Our objective is to study the entanglement built up between two free harmonic oscillators in such a spacetime due to their interaction through gravity. As before, we will assume that the two masses are comoving and have no peculiar velocity, \textit{i.e.} they are stationary with respect to each other and with respect to the cosmic fluid. This will let us invoke the non-relativistic approximation, as we had done earlier. 

We begin by writing the graviton field operators, linearized around the metric \eqref{dS}, schematically written as $g_{\mu\nu} = \eta_{\mu\nu}^{\rm dS} + h_{\mu\nu}$. For de Sitter, the vacuum state is much less ambiguous and one can write it down explicitly by requiring that it obeys the symmetries of the background and have the same short-distance behaviour as flat space (thereby, ruling out the so-called $\alpha$-vacua). The important thing to notice is that the Bunch-Davies mode functions, corresponding to the graviton field, can be written (in momentum space) as\footnote{We have introduced a traditional change of variables to the Mukhanov-Sasaki ones.}:
\begin{equation}\label{BD}
	f^{\rm BD}_k(\tau) = \frac{e^{-i k\tau}}{\sqrt{2k}}\left(1 - \frac{i}{k\tau}\right)\,.
\end{equation}
Expanding the off-shell graviton fields, responsible for entanglement between two stationary masses, requires additional details about polarization tensors and so on but those are unimportant for our purposes. Note that at early conformal times ($\tau \rightarrow -\infty$), the Bunch-Davies mode functions coincide with the Minkowksi vacuum since this corresponds to a given Fourier mode being well-inside the de Sitter horizon and hence unable to ``feel'' the spacetime curvature. On the other hand, the second term in \eqref{BD} plays the dominant role at late times ($\tau \rightarrow 0$), the so-called squeezing term, which is when the wavelength of the Fourier mode is bigger than the Hubble horizon. The reason behind reiterating these well-known facts shall be clear momentarily.

The effective potential, to leading order in the non-relativistic limit, between two point-like massive particles can be written as \cite{us, Araujo:2017clw}
\begin{eqnarray}\label{NewPotdS}
	V_{I}^{\rm dS}(\tau) = - G \, m_A \, m_B \left(\frac{1}{a\, |r_A - r_B|} +\frac{1}{4} H^2\, a\,|r_A - r_B|\right)\,,
\end{eqnarray}
where the scale factor $a = -1/(H\tau)$. This well-known result can also be derived numerically from the heavy mass (non-relativistic) limit of two-body scattering in de Sitter \cite{Ferrero:2021lhd}\footnote{Note that this effective potential is not the Newton-Hooke one. For why this is the case, see \cite{Araujo:2017clw}.}. As anticipated earlier for the general FLRW case, for small separation $a\, |r_A - r_B|\, H \ll 1$, that is when the two objects are well within the Hubble horizon, the standard Newtonian potential dominates as in flat space. The second term, which produces a constant \textit{repulsive} force, only kicks in when the two oscillators are farther apart from each other. In \cite{Ferrero:2021lhd}, numerical estimates show that the repulsive term in the effective potential starts dominating when $a\, |r_A - r_B|\, H \sim 0.5$. Thus, the gravitational interaction between the two massive objects pick up effects of spacetime curvature when the two bodies are distant enough to feel the horizon.

We can develop an intuition of how this potential comes about through a somewhat heuristic computation of using perturbation theory to compute the shift in the graviton vacuum due to the two masses \cite{Bose:2022uxe}. (We shall present the details of this elsewhere \cite{us}.) Importantly, note that the only difference is that we use the Bunch-Davies modes from \eqref{BD} instead of the Minkowski ones used in \cite{Bose:2022uxe}. The final result can be quoted in the form of two integrals, the first one 
$$\int \frac{d^3 k}{(2\pi)^3} \text{ }\frac{1}{\textbf{k}^2}\left\{ e^{i \textbf{k}\cdot (\textbf{r}_A - \textbf{r}_B)} + e^{-i \textbf{k}\cdot (\textbf{r}_A - \textbf{r}_B)}\right\}$$
yields the standard $1/|r_A - r_B|$ term while a second integral
$$\frac{1}{\tau^2}\int \frac{d^3k}{(2\pi)^3}\text{ }\frac{1}{\textbf{k}^4}\left\{ e^{i \textbf{k}\cdot (\textbf{r}_A - \textbf{r}_B)} + e^{-i \textbf{k}\cdot (\textbf{r}_A - \textbf{r}_B)}\right\}$$ 
gives the new term $\propto |r_A - r_B|$. As is usual, one discards the self-energy terms in these computations and only keeps the infrared terms responsible for the effective potential.

The reason to emphasize this detail is to say that the ``squeezing'' term in \eqref{BD}, corresponding to the second integral above, is responsible for the new term in \eqref{NewPotdS}. In other words, the same term which is responsible for producing a scale-invariant power spectrum of cosmological observables is the one behind producing the correction to the non-relativistic gravitational potential. Why does this matter? On looking closely at \eqref{NewPotdS}, the reader will immediately realize that although the effective potential is modified, \textit{the entanglement profile between the two oscillators in de Sitter will remain exactly the same as in Minkowski space}. This is because the new term is a bi-local one and cannot lead to entanglement between the two masses. This is quite a remarkable finding -- two massive objects in de Sitter will be entangled in the same manner as in flat space, to leading order in the non-relativistic approximation, even when pushed far enough to be able to feel the effects of the horizon. And it immediately tells us, without doing any computation, that a general FLRW solution will produce a different entanglement profile between two point masses. In other words, if the background is not de Sitter, then the mode functions for the graviton will not have the same ``squeezing'' term (we know this since the scale invariance is a specific property of de Sitter and not of general FLRW metrics) and the new contribution to the interaction potential, due to spacetime curvature, will generically lead to a different entanglement profile. \ul{More generally, however, we have shown that, in principle, gravity-induced entanglement between two massive objects (confined within two harmonic traps) is dependent on the curvature of the background cosmology and is, in particular, sensitive to it.}

Finally, let us end with a discussion regarding the practical utility of our thought experiment. Even for flat space, our abilities to prepare macroscopic objects in spatial superpositions is not yet advanced enough to test the entanglement produced by gravitons. Nevertheless, we have come a long way from the days of the Chapel Hill Conference to be ambitious enough to implement this thought experiment in a lab. A nice analogy would be how our ancestors set sail in ships across the world to demonstrate that the Earth is a sphere. (Similar ideas regarding the topology of the Universe are being explored these days \cite{Yashar}.) We are now at that point when we can devise of an idealized thought experiment to show how entanglement due to the gravitational field contains signatures of the spacetime curvature of the background. 

Can we then hope to capture any of the curvature effects discussed in this essay, perhaps in the not too distant future?  The first thing to do would be a fully relativistic treatment of this phenomenon, which was avoided in this essay, as it adds technical complications and takes away from the main physics of our argument. Looking ahead, we might then be able to capture the graviton entanglement between two CMB photons which travel over vast cosmological distances, through different expansion eras, since such photons travel fairly unhindered and might not have suffered decoherence. Even if the entanglement due to gravity is very weak, being able to travel over such large distances might add up small effects and result in something that is detectable. Moreover, our thought experiment is, of course, not confined to cosmology and it is reasonable to think that repeating this exercise with two masses, say on the surface of the moon and the Earth, might capture some facets of the Schwarzschild metric around the Earth, requiring us to go beyond flat space results. Since any detection of gravity-mediated entanglement would be a first confirmation of perturbative quantum gravity, all possible routes need to be explored in this direction.  

In summary, we have conclusively shown that gravity-induced entanglement depends on the spacetime curvature and can, at least in principle, be thought of as an alternate probe for the latter.

%%%%%%%%%%%%%%%%%%%%%%%%%%%%%%%%%%%%%%%%%%%%%%%%%%

\vspace{3mm}
\noindent {\bf Acknowledgements:} SB am grateful to B\'eatrice Bonga, Jaime Calder\'on-Figueroa, Anton Ilderton and Donal O'Connell for discussions. SB is supported in part by the Higgs Fellowship and by the STFC Consolidated Grant ``Particle Physics at the Higgs Centre''.

%%%%%%%%%%%%%%%%%%%%%%%%%%%%%%%%%%%%%%%%%%%%%%%%%%%%

\end{document}